      [1999/12/01 v1.4c Il Nuovo Cimento]
\documentclass{cimento}

\usepackage{graphicx}  
\title{In-situ Particle Acceleration in Collisionless Shocks}
\author{C. B.~Hededal\from{ins:c}\ETC,
T.~Haugb\o{}lle\from{ins:c}, J. T.~Frederiksen\from{ins:s}
        \atque
\AA{}.~Nordlund\from{ins:c}} \instlist{\inst{ins:c} The Niels Bohr
Institute, Department of Astrophysics, Juliane Maries Vej 30, 2100
Copenhagen, Denmark
  \inst{ins:s} Stockholm Observatory, Roslagstullbacken 21, 106 91 Stockholm, Sweden}
\PACSes{\PACSit{00.00}{By the way, which PACS is it, the 00.00?
GOK.} \PACSit{---.---}{\ldots}}
\begin{document}

\maketitle

\begin{abstract}
The outflows from gamma ray bursts, active galactic nuclei and
relativistic jets in general interact with the surrounding media
through collisionless shocks. With three dimensional relativistic
particle--in--cell simulations we investigate such
shocks. The results from these experiments show that small--scale
magnetic filaments with strengths of up to percents of equipartition
are generated and that electrons are accelerated to power law
distributions $N(\gamma)\propto{}\gamma^{-p}$ in the vicinity of the filaments
through a new acceleration mechanism. The acceleration is locally
confined, instantaneous and differs from recursive acceleration
processes such as Fermi acceleration. We find that the proposed
acceleration mechanism competes with thermalization and becomes
important at high Lorentz factors.
\end{abstract}

\section{Introduction}
Observations of astrophysical objects with relativistic plasma
outflows generally show frequency spectra dominated by power-law
segments. These objects count gamma ray bursts (GRBs) and active
galactic nuclei. The general interpretations of such observations
are that 1) electrons are accelerated to power-law distributions
$N(\gamma)\propto{}\gamma^{-p}$ in the collisionless shock interface
between the relativistic ejecta and the surrounding media and 2) the
accelerated electrons radiate through synchrotron radiation in the
downstream region of the shock. The magnetic field strength required
to account for the observed radiation spectra is typically of the
order of percents of equipartition.
A complete physical model for relativistic collisionless shocks should
account for both the origin of such strong magnetic fields and for
acceleration of electrons to power-law distributions.

Regarding the origin of the magnetic field, particle-in-cell (PIC)
simulations have shown that the Weibel two-stream instability can
generate a small scale, but strong, magnetic field transverse to the
jet flow ($\epsilon_B\simeq0.05$ where $\epsilon_B$ describes the
fraction of total injected kinetic energy that is converted to
magnetic energy)
\cite{bib:frederiksen2004,bib:kazimura1998,bib:medvedev1999,bib:nishikawa2004,bib:silva2003}.
Other PIC simulations have found that the ambient large scale magnetic
field neccesary to quench this instability has to be significantly stronger
than the $\mu$G field typically found in the interstellar medium (ISM)
\cite{bib:hededal2004b,bib:sakai2004}.

Regarding the non-thermal electron acceleration, Monte Carlo
test-particle simulations have shown that Fermi acceleration can
provide electron power-law distributions under some assumptions
about the shock and magnetic field \cite{bib:niemiec}. This
mechanism has, however, not been conclusively demonstrated to occur
in ab initio particle simulations. PIC simulations have shown that
both thermal and non-thermal particle acceleration take place in the
shock transition region
\cite{bib:hededal2004a,bib:hoshino2002,bib:nishikawa2004,bib:saito2004}.

In this paper, we discuss the particle dynamics in collisionless
shocks based on results from a three dimensional relativistic PIC
code. We propose a new particle acceleration mechanism that is
itself a consequence of the Weibel two-stream instability
\cite{bib:hededal2004a}.

\section{Numerical Model and Results}
We have performed computer experiments using a charged
particle-in-cell code. The code works from first principles by
solving Maxwell's equations for the electromagnetic fields and the
Lorentz force equation of motion for the particles. The fields are
defined on a numerical grid and the particles are defined in a
continuous phase space. We use a computational box with
$125\times125\times2000$ grid points and a plasma consisting of
$8\times10^8$ ions and electrons. The ion rest-frame plasma
frequency is $\omega_{pi}=0.075$, rendering the box 150 ion skin
depths long. The electron rest-frame plasma frequency is
$\omega_{pe}=0.3$ in order to resolve the microphysics of the
electrons. Hence, the ion-to-electron mass ratio is $m_i/m_e = 16$.

\begin{figure}[!t]
\includegraphics[width=0.5\textwidth]{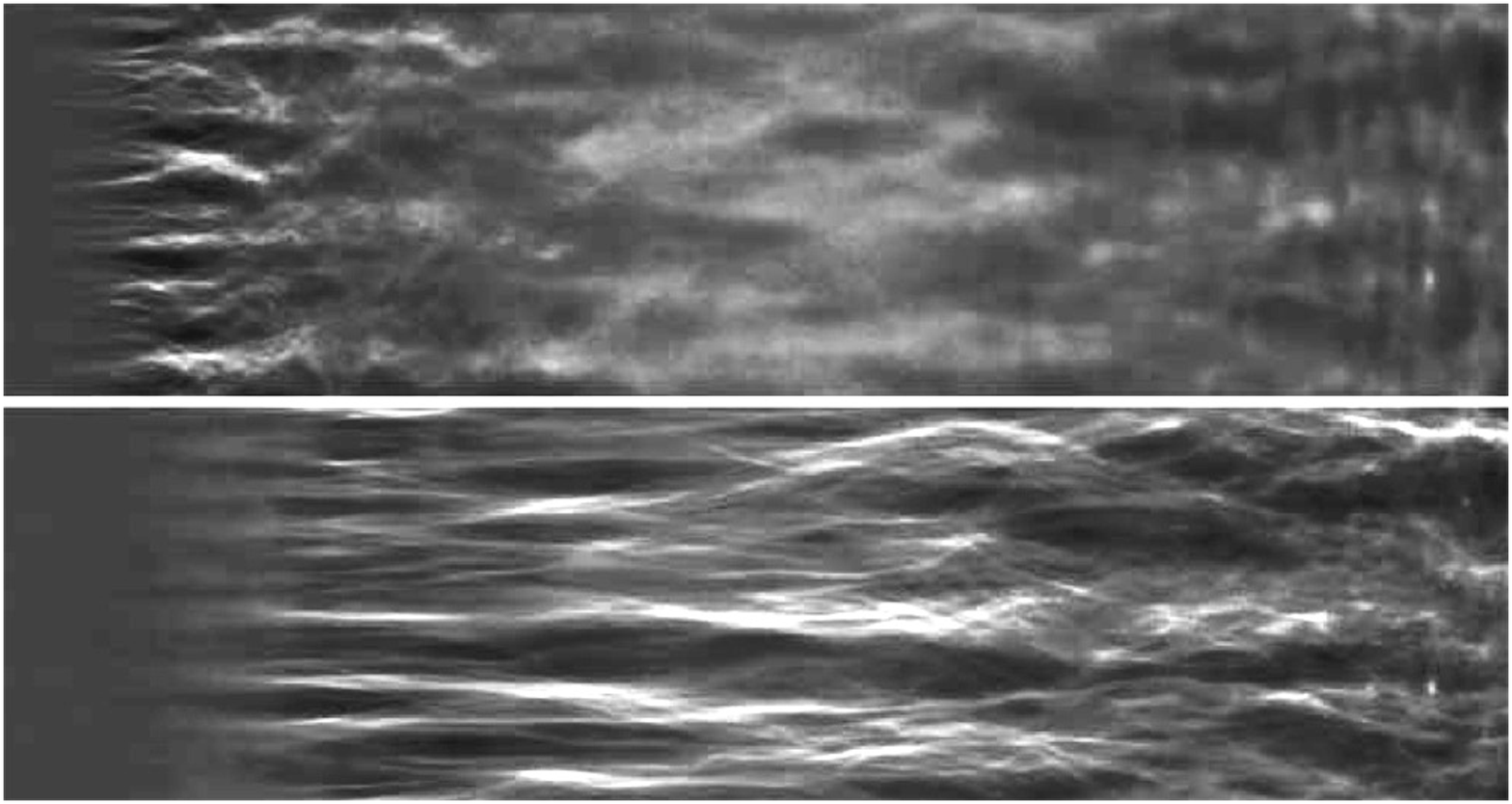}     
\includegraphics[width=0.5\textwidth]{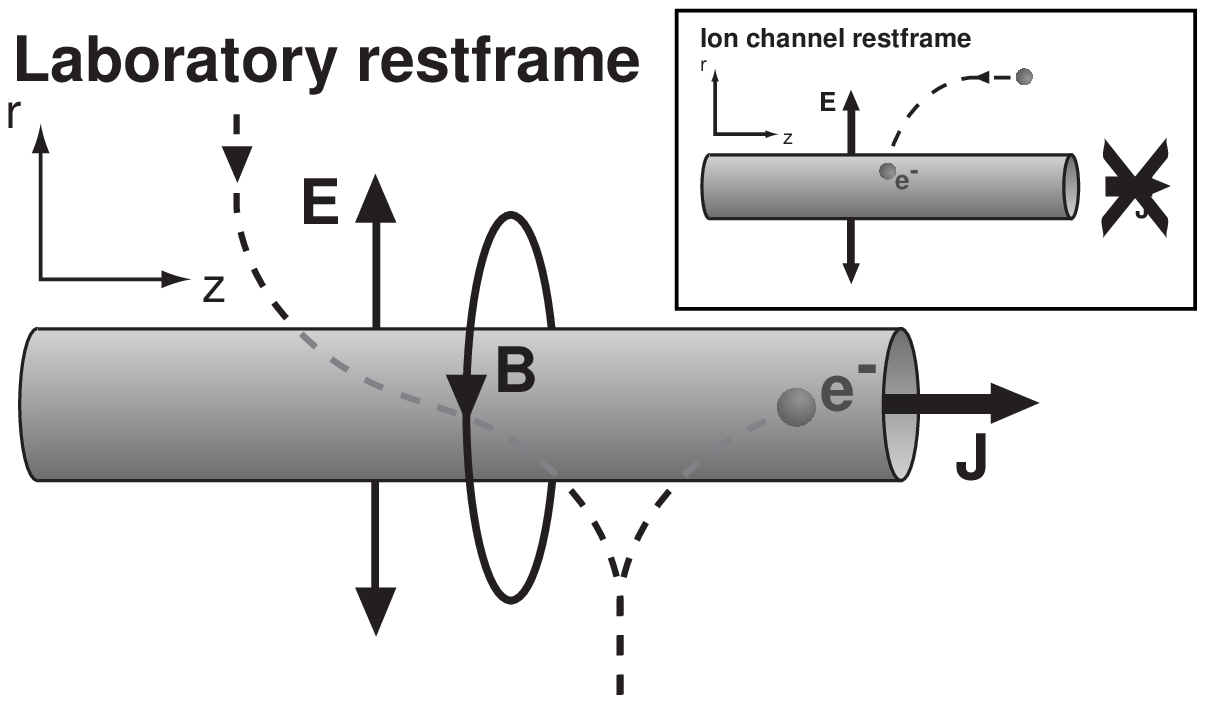}     
\caption{{\em Left panel:} The Weibel two-stream instability forms
current channels. The electrons ({\em top panel}) undergo the
instability faster the the heavier ions ({\em bottom panel}). {\em
Right panel:} A schematic view of an ion current channel surrounded
by an electric and a magnetic field. Electrons in the vicinity of
the current channels are subject to a Lorentz force with both an
electric and magnetic component, working together to accelerate the
electrons along the ion flow. Crossing the center of the channel,
the process reverses, leading to an oscillating movement along the
channel.}\label{fig:current}
\end{figure}

In the experiments we study the microphysics of two colliding plasma
populations. The experiments are carried out in the rest frame of
one of the populations (downstream, e.g., a jet). In this frame, a
less dense population (upstream, e.g. the ISM) is continuously
injected at the leftmost boundary $z=0$. We have performed two runs
with relativistic velocities corresponding to Lorentz factors
$\Gamma=3$ and $\Gamma=15$. Initially, the jet plasma is denser by a
factor of three and both plasma populations are unmagnetized. The
boundaries in the direction transverse to the jet flow ($x, y$) are
periodic. At the $z$-boundaries, electromagnetic waves are absorbed
and we allow particles to escape in order to avoid unphysical
feedback. The experiment last until $t_{max} =$ 340
$\omega_{pi}^{-1}$, which is sufficient for the continuously
injected upstream ISM-plasma to travel
2.3 times the length of the box.

\begin{figure}[!t]
\includegraphics[width=0.5\textwidth]{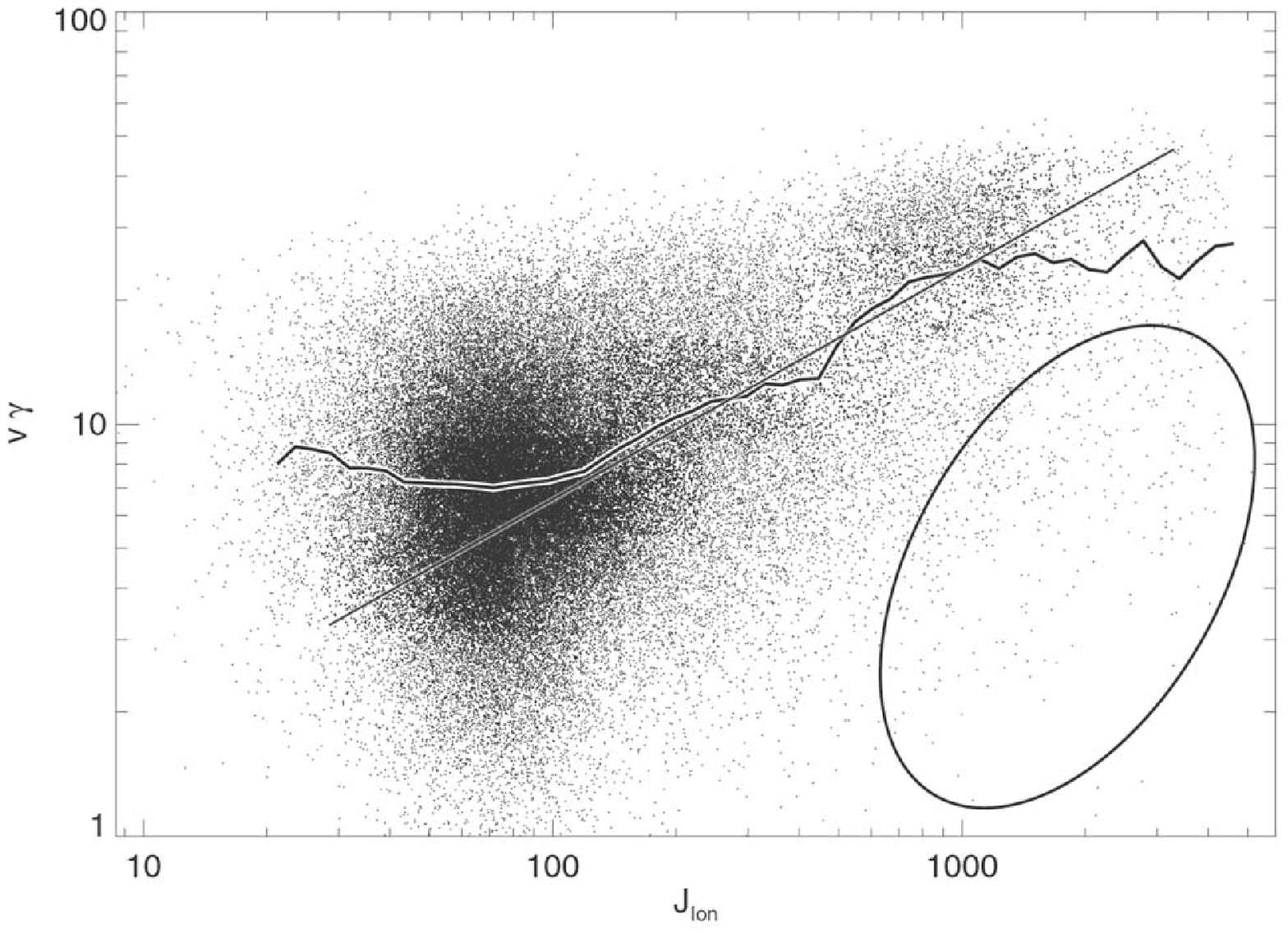}     
\includegraphics[width=0.5\textwidth]{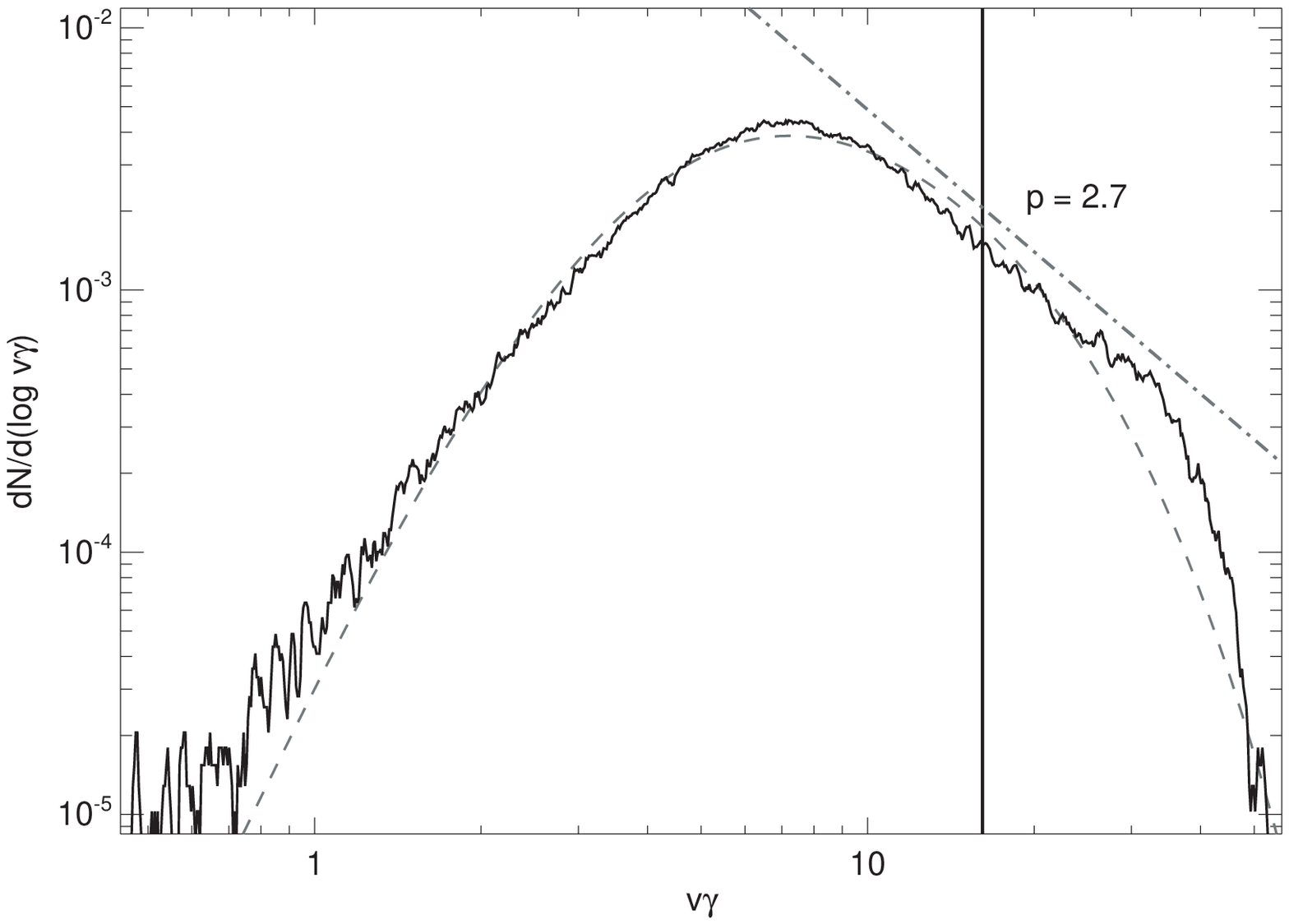}     
\caption{{\em Left panel:} A scatter plot of the local ion current
density $J_{ion}$ versus the four velocity of the electrons in a
region downstream of the shock. Overplotted is a line (thin) showing
the average four velocity as a function of $J_{ion}$, and a line
(thick) showing a straight line fit. {\em Right panel:} The
normalized electron particle distribution function downstream of the
shock. The dot--dashed line is a power law fit to the non--thermal
high energy tail, while the dashed curve is a Lorentz-boosted
thermal electron population. The horizontal line at $v\gamma=15$
indicates the plasma injection velocity.}\label{fig:powerlaw}
\end{figure}

The results from the experiments show how the Weibel two-stream
instability forms current filaments in the region where the two
plasma populations interpenetrate and how these filaments merge into
increasingly larger patterns further downstream (Fig.
\ref{fig:current} {\em left panel}) (see also
\cite{bib:frederiksen2004,bib:medvedev2005,bib:nishikawa2004,bib:silva2003}).
The induced magnetic field grows to $\epsilon_B\simeq1\%$. We identify
two mechanism that accelerate electrons. The first is thermalization
caused by random deflections by the generated small scale
electromagnetic field. This mechanism is dominant for the $\Gamma=3$
run. In the $\Gamma=15$ run a second acceleration mechanism becomes important.
Here we briefly discuss this mechanism (for details please see \cite{bib:hededal2004a}):

The generated ion current channels are Debye shielded by the heated electrons. At
distances less than the Debye length, the current channels are
surrounded by transverse electric fields that accelerate the
electrons toward the current channels. The acceleration happens in
an approximately electrostatic field and is a simple consequence of
potential energy being converted into kinetic energy. Therefore, the
electrons are decelerated again when leaving the current channel and
reach their maximal velocities at the centers of the current
channels (Fig. \ref{fig:current}). It has been shown that a spatial
Fourier decomposition of the transverse ion filaments in the box
exhibits power-law behavior
\cite{bib:frederiksen2004,bib:medvedev2005}. Hence, the number of
accelerated electrons is expected to reflect this power-law
behavior. This relation is confirmed in Fig. \ref{fig:powerlaw}. The
left panel shows a scatter plot of electron momentum ($v\gamma$)
plotted against the strength of the electric current sampled at each
electrons position. There is a power-law correlation between the
electron energy and the local current density. Some deviations from
the power-law is found: 'Cold' trapped thermal electrons (indicated
with the ellipse) exist inside the ion current channel and count
towards lowering the average four velocity at high $J_{ion}$. Also,
heated electrons are found outside the current channels which
increase the average four velocity at low $J_{ion}$. The electron
distribution function can be seen in the right panel of Fig.
\ref{fig:powerlaw}. It show a thermal distribution with a power-law
component at high energies. The slope of the power-law segment
corresponds to a spectral index of p=2.7.

\section{Conclusions}
We have investigated magnetic field generation and
particle acceleration in collisionless shocks using
three dimensional charged particle-in-cell experiments.
The results are
applicable to interactions between relativistic outflows and the
interstellar medium. Such relativistic outflows occur in GRBs and in
jets from compact objects.

We observe how the Weibel two-stream instability is excited and find
that the non-linear stage of the instability penetrates at least 150
ion skin depths into the shock ramp. Extrapolating to the real
electron-ion mass ratio, this is more than 6000 electron skin
depths. The instability efficiently creates a strong, small-scale
magnetic field ($\epsilon_B\simeq1\%$).

Regarding particle acceleration, the results show that thermalization
dominates the electron distribution in the simulation run
where a jet is expanding with $\Gamma=3$. At $\Gamma=15$, a new
acceleration mechanism becomes important in the high energy tail of
the spectrum. The acceleration is non-thermal and the resulting
electron distribution function has a power-law segment. The slope of
this segment corresponds to $p=2.7$.

We emphasize that the generation of magnetic field and the
non-thermal acceleration of electrons are two highly interconnected
processes in collisionless shocks. The acceleration is closely
tied to the spatial distribution of electric current filaments.
Further large scale simulations are need to determine how this slope
depends on different parameters.

When the injected ISM ions has traveled through the simulation box
they are not merged with the jet ions to a single population. This
indicates that larger simulations are needed to cover the whole
shock ramp.

\acknowledgments We would like to thank the Danish Center for
Scientific Computing for granting the computer resources that made
this work possible.

\end{document}